\documentclass[aps,prl,twocolumn,groupedaddress,showkeys,superscriptaddress]{revtex4-1}
\usepackage{graphicx}           
\usepackage{amsmath}
\usepackage{amssymb}            
\usepackage{upgreek}
\usepackage{braket}             
\usepackage{hyperref}
\usepackage{xspace} 
\usepackage[table,usenames,dvipsnames]{xcolor}
\usepackage{colortbl}
\usepackage{verbatim}

\newcommand{\mus}{\,\ensuremath{\mathrm{\upmu s}}\xspace}

\newcommand{\mev}{\,\ensuremath{\mathrm{meV}}\xspace}

\newcommand{\Mev}{\,\ensuremath{\mathrm{MeV}}\xspace}
\newcommand{\ghz}{\,\ensuremath{\mathrm{GHz}}\xspace}
\newcommand{\thz}{\,\ensuremath{\mathrm{THz}}\xspace}
\newcommand{\fm}{\,\ensuremath{\mathrm{fm}}\xspace}

\newcommand{\Het}{\ensuremath{^3}{\rm He}\xspace}

\newcommand{\rh}{\ensuremath{r_h}}
\newcommand{\ra}{\ensuremath{r_\alpha}}

\newcommand{\twoS}{\ensuremath{\mathrm{2S}}}
\newcommand{\twoP}{\ensuremath{\mathrm{2P}}}
\newcommand{\twoStwoP}{$\mathrm{2S\rightarrow 2P}$}
\newcommand{\twoPoneS}{$\mathrm{2P\rightarrow 1S}$}

\begin{document}

\title{The helion charge radius from laser spectroscopy of muonic helium-3 ions}

\newcommand\MPQ{Max-Planck-Institut f\"ur Quantenoptik, 85748 Garching, Germany}
\newcommand\JGU{Johannes Gutenberg Universit\"at Mainz, Institut f\"ur Physik, QUANTUM, \& Exzellenzcluster PRISMA$^+$, 55099 Mainz, Germany}
\newcommand\ETH{Institute for Particle Physics and Astrophysics, ETH Zurich, 8093 Zurich, Switzerland}
\newcommand\PSI{Paul Scherrer Institute, 5232 Villigen, Switzerland}
\newcommand\IFS{Institut f\"ur Strahlwerkzeuge, Universit\"at Stuttgart, 70569 Stuttgart, Germany}
\newcommand\COIMBRA{LIBPhys, Physics Department, Universidade de Coimbra, \mbox{P-3004-516 Coimbra, Portugal}}
\newcommand\LISBOA{Laborat\'{o}rio de Instrumenta\c{c}\~{a}o, Engenharia Biom\'{e}dica e F{\'\i}sica da Radia\c{c}\~{a}o (LIBPhys-UNL) e Departamento de F{\'\i}sica da Faculdade de Ci\^encias e Tecnologia da Universidade Nova de Lisboa, Monte da Caparica, 2892-516 Caparica, Portugal}
\newcommand\LKB{Laboratoire Kastler Brossel, UPMC-Sorbonne Universit\'{e}s, CNRS, ENS-PSL Research University, Coll\`{e}ge de France, 4 place Jussieu, case 74, 75005 Paris, France}
\newcommand\TAIWAN{Physics Department, National Tsing Hua University, Hsincho 300, Taiwan}
\newcommand\AVEIRO{i3N, Universidade de Aveiro, Campus de Santiago, 3810-193 Aveiro, Portugal}
\author{Karsten Schuhmann}\affiliation{\ETH}
\author{Luis M.~P.\ Fernandes}\affiliation{\COIMBRA}
\author{Fran\c{c}ois Nez}\affiliation{\LKB}
\author{Marwan Abdou Ahmed}\affiliation{\IFS}
\author{Fernando D.\ Amaro}\affiliation{\COIMBRA}
\author{Pedro Amaro}\affiliation{\LISBOA}
\author{Fran\c{c}ois Biraben}\affiliation{\LKB}
\author{Tzu-Ling Chen}\affiliation{\TAIWAN}
\author{Daniel S.\ Covita}\affiliation{\AVEIRO}
\author{Andreas J.\ Dax}\affiliation{\PSI}
\author{Marc Diepold}\affiliation{\MPQ}
\author{Beatrice Franke}\affiliation{\MPQ}
\author{Sandrine Galtier}\affiliation{\LKB} 
\author{Andrea L.\ Gouvea}\affiliation{\COIMBRA}
\author{Johannes G\"otzfried}\affiliation{\MPQ}
\author{Thomas Graf}\affiliation{\IFS}
\author{Theodor~W.\ H\"ansch}\affiliation{\MPQ}
\author{Malte Hildebrandt}\affiliation{\PSI}
\author{Paul Indelicato}\affiliation{\LKB}
\author{Lucile Julien}\affiliation{\LKB}
\author{Klaus Kirch}\affiliation{\ETH}\affiliation{\PSI}
\author{Andreas Knecht}\affiliation{\PSI}
\author{Franz Kottmann}\affiliation{\ETH}\affiliation{\PSI}
\author{Julian J.\ Krauth}\affiliation{\MPQ}\affiliation{\JGU}
\author{Yi-Wei Liu}\affiliation{\TAIWAN}
\author{Jorge Machado}\affiliation{\LISBOA}
\author{Cristina M.~B.\ Monteiro}\affiliation{\COIMBRA}
\author{Fran\c{c}oise Mulhauser}\affiliation{\MPQ}
\author{Boris Naar}\affiliation{\ETH}
\author{Tobias Nebel}\affiliation{\MPQ}
\author{Joaquim M.~F.\ dos Santos}\affiliation{\COIMBRA}
\author{Jos\'{e} Paulo Santos}\affiliation{\LISBOA}
\author{Csilla I.\ Szabo}\affiliation{\LKB}
\author{David Taqqu}\affiliation{\ETH}\affiliation{\PSI}
\author{Jo\~{a}o F.~C.~A.\ Veloso}\affiliation{\AVEIRO}
\author{Andreas Voss}\affiliation{\IFS}
\author{Birgit Weichelt}\affiliation{\IFS}
\author{Aldo Antognini}\email[]{aldo.antognini@psi.ch}\affiliation{\ETH}\affiliation{\PSI}
\author{Randolf Pohl}\email[]{pohl@uni-mainz.de}\affiliation{\MPQ}\affiliation{\JGU}

\collaboration{The CREMA Collaboration}
\noaffiliation

\date{\today}

\begin{abstract}

Hydrogen-like light muonic ions, in which one negative muon replaces all the electrons,
are extremely sensitive probes of nuclear structure, because the large muon
mass increases tremendously the wave function overlap with the nucleus.
Using pulsed laser spectroscopy we have measured three 2S-2P transitions in the  muonic helium-3 ion ($\mu^3$He$^+$), an ion formed by a negative muon and bare helium-3 nucleus.
This allowed us to extract the Lamb shift 
$E(2P_{1/2}-2S_{1/2})= 1258.598(48)^{\rm exp}(3)^{\rm theo}$\mev,
the 2P fine structure splitting
$E_{\rm FS}^{\rm exp} = 144.958(114)$\mev,
and the 2S-hyperfine splitting (HFS)
$E_{\rm HFS}^{\rm exp} = -166.495(104)^{\rm exp}(3)^{\rm theo}$ \mev
in $\mu^3$He$^+$.
Comparing these measurements to theory  we
determine the rms charge radius of the helion ($^3$He nucleus) to be \rh =
1.97007(94)\fm. This radius represents a benchmark for few nucleon theories and opens the way for precision tests in $^3$He atoms and $^3$He-ions. 
This radius is in good agreement with the 
value from elastic electron scattering, but a factor 15 more accurate.
Combining our Lamb shift measurement with our earlier one in $\mu^4$He$^+$
we obtain  
  $\rh^2-\ra^2 = 1.0636(6)^{\rm exp}(30)^{\rm theo}\fm^2$
to be compared to results from the isotope shift measurements in regular He atoms, which are however
affected by long-standing tensions.
By comparing $E_{\rm HFS}^{\rm exp}$ with theory we also obtain the two-photon-exchange contribution (including higher orders) which is another important benchmark for ab-initio few-nucleon theories
aiming at understanding the magnetic and current structure of light nuclei.

\end{abstract}


\maketitle

\section{Introduction}

Precise and accurate values of nuclear radii are necessary for advancing the theory prediction in simple atomic and molecular systems, that could be exploited to test bound-state Quantum Electro Dynamics (QED) predictions, to determine fundamental constants such as the Rydberg constant and the electron mass and to search for new physics~\cite{Antognini:2022xoo}. 
Moreover these radii represent rigorous benchmarks for nuclear structure theory with the peculiar  enhanced sensitivity to the long range behaviour of the nuclear wave functions.
Spectacular advances have been obtained 
recently  for nucleon-nucleon currents and form factors of light nuclei  using nuclear interactions derived from chiral effective field theory  which can be systematically improved~\cite{Ji:2018:nuclPOL, Epelbaum:2019jbv, Piarulli:2012bn}. 
On the experimental side, laser spectroscopy of muonic atoms with their exquisite sensitivity to nuclear properties has been recently established~\cite{Antognini:2013:Science_mup2, Pohl:2016:mud, Krauth:2021foz}.
Owing to the 200 times larger muon mass compared to the electron mass, these muonic atoms (ions) where a single negative muon is orbiting a bare nucleus,
have a dramatically enhanced sensitivity to the nuclear charge radii compared to regular atoms.
This sensitivity arises from the enhanced overlap of the atomic wavefunction with the nucleus which scales with the  third power of the orbiting particle mass (reduced mass). Indeed the leading-order finite-nuclear-size effect takes the form 
\begin{align}
\Delta E_\mathrm{FNS} (n,l) &= \frac{2}{3n^3}(Z\alpha)^4m_r^3r^2 \delta_{l0}\, ,
\label{eq:finite-size}\
\end{align}
where $n$ is the principal quantum number, $l$ the angular momentum, $\alpha$ the fine structure constant, $Z$ the atomic number, $m_r$ the  reduced mass,  and $r$ the nuclear charge radius.
The Kronecker $\delta_{l0}$ indicates that only S-states energy levels are  affected in leading approximation by the finite size due to their overlap with the nucleus.

\begin{figure}
  \includegraphics[width=0.65\linewidth]{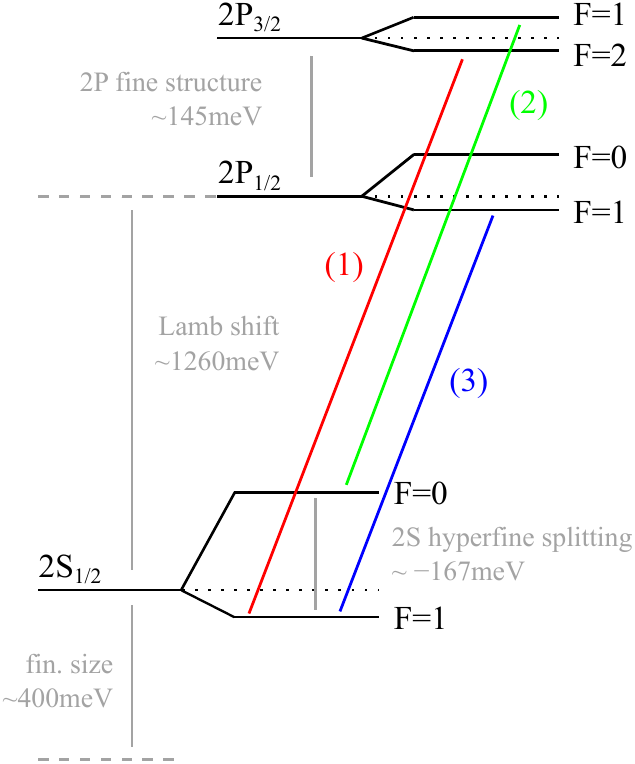}
  \caption{Scheme (not to scale) of the $n=2$ energy levels in $\mu^3$He$^+$ and
    the measured transitions. Due to the negative magnetic moment of the
    helion, the ordering of the hyperfine levels is reversed.
    }
  \label{fig:levels}
\end{figure}

In this work, we present the measurement of three \twoStwoP{}
transitions of the muonic helium-3 ion $\mu^3$He$^+$ (a two-body ion formed by a negative muon and a bare $^3$He nucleus) as shown in Fig.\,\ref{fig:levels}
from which we extracted the 2S-2P Lamb shift, 
the 2S hyperfine splitting 
and the 2P fine splitting.
From the Lamb shift  we have then extracted the rms charge radius of the helion \rh{} with an unprecedented relative precision of $7\times10^{-4}$, improving the previous best value \cite{Sick:2014:HeZemach} from elastic electron scattering by a factor of 15. 
From the comparison between  2S-HFS and theory, we  extracted the  two-photon-exchange contribution which is the leading-order nuclear structure dependent contribution for the HFS.
Hence, these measurements  provide important benchmarks for nuclear theories, 
and pave the way for advancing precision tests of two- and three-body QED when combined with ongoing efforts in regular He atoms~\cite{Shiner:1995:heliumSpec,Rooij:2011:HeSpectroscopy,    CancioPastor:2012:PRL108,Patkos:2016:HeIso,Zheng:2017:He_Iso,Rengelink:2018:4He, PhysRevA.101.062507, vanderWerf:2023cuv} and He$^+$ ions~\cite{Moreno:2023amv, Krauth:2019:HePlus}.

The 2S-2P transition frequencies were measured by pulsed laser spectroscopy at wavelengths around 850-940~nm (frequencies of 310-350~THz) to an accuracy of about 20~GHz corresponding to relative accuracies of about 50~ppm. 
The resonances were exposed by detecting the $K_\alpha$ X-ray of 8~keV energy
emitted from \twoPoneS{}  de-excitation following a successful laser transition from the 2S to the 2P state.
The 50~ppm measurement precision has to be compared to the energy shift caused by the finite-size effect (see Eq.~\eqref{eq:finite-size}) that contributes as much as 25\% to the 2S-2P energy splitting  owing to its $m_r^3Z^4$ dependence.
The binding energy of this hydrogen-like system, scaling as $Z^2m_r$, is strongly enhanced compared to hydrogen, while the atomic size (Bohr radius) scaling as $1/Zm_r$ is strongly reduced making this atom immune to external perturbation.
Because of its $m_rZ^4$ dependence, the decay rate from the 2P state is also vastly increased  resulting in a 2P-linewidth of 319~GHz. This broad linewidth   
represents by far the main limitation to our experimental precision.
Having in mind these energy scales, sensitivity to nuclear properties and immunity to external perturbations helps framing the requirements for the spectroscopy experiment and understanding of the experimental setup.

\section{Principle and experimental setup}

$\mu^3$He$^+$ ions are formed in highly excited states by stopping a keV-energy  muon in a low-pressure (2 mbar) $^3$He gas target  at room temperature and placed in a 5~T solenoid.
The newly formed $\mu^3$He$^+$ ions  deexcite to the 1S state in a fast (ns-scale) and complex cascade process with a fraction of about 1\% reaching the metastable 2S state whose lifetime is approximately 1.7~$\mu$s at this target pressure.
This lifetime is sufficiently long to enable pulsed laser spectroscopy of the 2S-2P splitting.
For this purpose the muon entering the target is being detected to trigger a laser system that delivers a pulse to excite the 2S into the  2P state after a delay of about 1\,\mus{}.
The laser pulse is injected into a multipass cell formed by two elongated mirrors that fold the light back and forth to illuminate the elongated muon stopping volume.
A successful laser excitation is established by detecting the  8~keV energy $K_\alpha$ X-ray from the 2P  de-excitation into the ground state. 
This is accomplished  using  two rows of large area avalanche photodiodes (LAAPDs) placed above and below the muon stopping volume, respectively and covering 30\% solid angle.
The 2S-2P resonances are eventually obtained by plotting  the number of 
$K_\alpha$ X-rays detected in time  coincidence with the laser light as a function of the laser frequency. 

\begin{figure}
  \includegraphics[width=\linewidth]{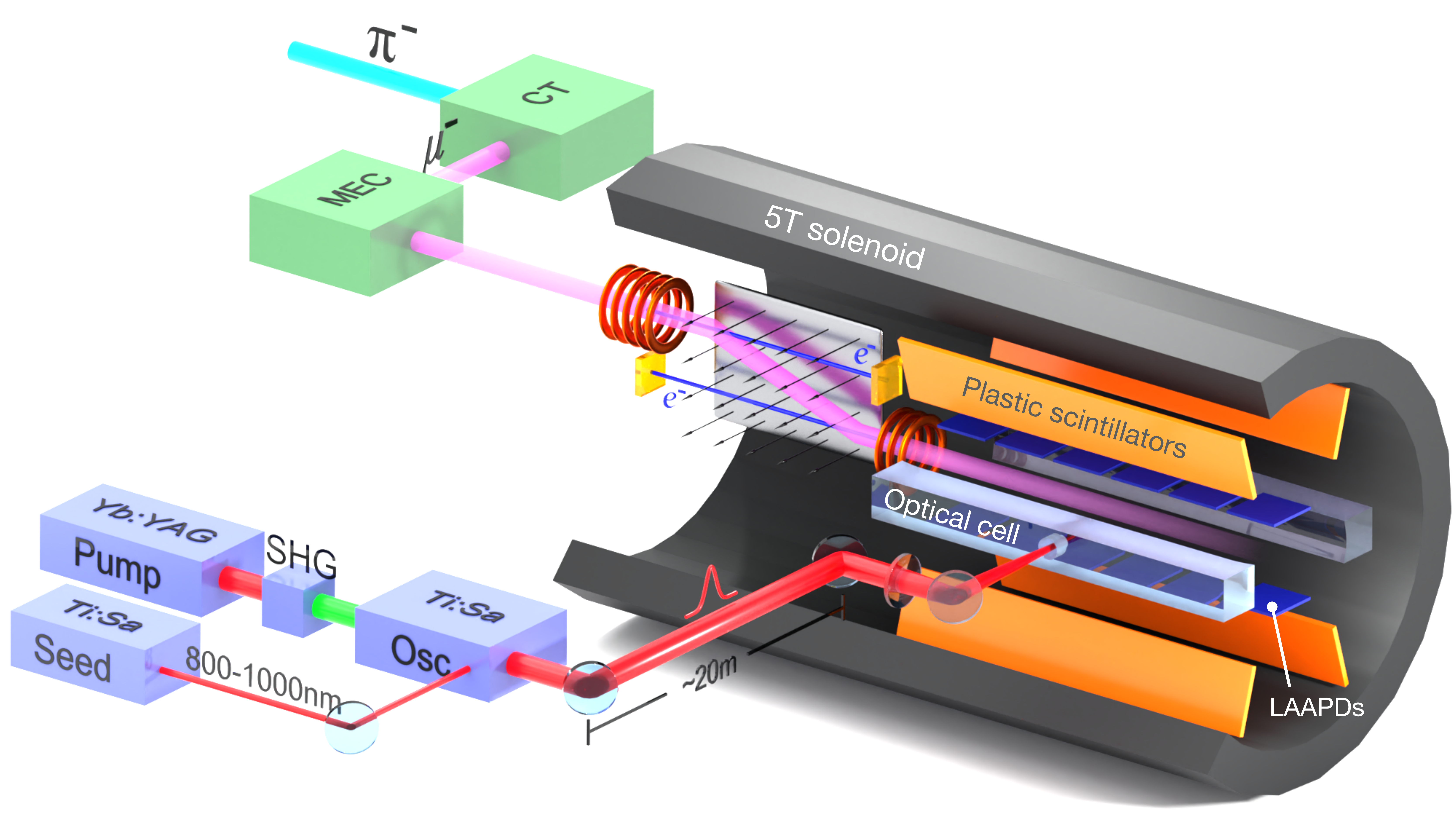}
  \caption{Sketch of the experimental setup.  CT cyclotron trap, MEC muon extraction channel, Ti:Sa Titanium-sapphire laser, Yb:YAG thin-disk laser, SHG second harmonic generation. The arrows indicate the electric field of the $E\times B$ Wien-filter. }
  \label{fig:setup}
\end{figure}

A sketch of the setup is shown in Fig.\,\ref{fig:setup}. The experiment has been performed at the $\pi$E5 beamline of the CHRISP facility at the Paul Scherrer Institute (PSI), Switzerland. 
Here, $102\Mev/c$ negative pions
are injected tangentially into 
a cyclotron trap (CT) formed by two coils generating a B-field with  a magnetic-bottle geometry
\cite{Gotta:2021vke}. 
A fraction of the muons from the pion decays are trapped in the B-field of the CT.
Passing multiple times a thin foil placed in the trap mid-plane, these muons are decelerated down to  20-40 keV energy. 
At this low energy, the electric field which is applied along the trap axis imparts sufficient longitudinal momentum to the muons so that they can escape the magnetic confinement of the CT in axial direction.
The escaping muons are coupled into a toroidal magnetic field,  the muon extraction channel (MEC),  that acts  as a momentum filter. 
The MEC transports the muons into a region of lower background where the spectroscopy experiment can be performed shielded from the large background of neutrons, gammas and other charged particles present in the CT.
From the MEC toroidal field of 0.1~T, the muon beam is focused into the 5~T solenoid.
Before entering the He target with an average rate of about 500\,s$^{-1}$,
the muons are passing two stacks of few $\mu$g/cm$^2$ thin carbon  foils placed at high voltages that serve two purposes: firstly they reduce the muon energy to few keV while obtaining some frictional cooling~\cite{Muehlbauer:1999:FrictCool}, and secondly they generate a muon entrance signal to trigger the laser system.
The muon entrance signal is obtained by detecting the secondary electrons ejected from the carbon foils by the passing muon. 
The foils are biased to accelerate the electrons  towards two plastic scintillators coupled to photomultipliers.
An $E \times B$ Wien  filter  separates the muon from the  electrons generated in the first stack of foils.
To increase the laser trigger quality, i.e., to fire the laser only for muons which have maximal probability to stop in the He gas target, a fine selection of the muon momentum is performed.
This is achieved requiring a coincidence with the optimal time-of-flight between the two electron signals from the two carbon stacks.
The muons then pass a thin (2~$\mu$g/cm$^2$) entrance window of Formvar before slowing down in the low-density  He gas  giving rise to a 20~cm long stopping distribution.
The gas in the target is kept clean by a circulation system with  a cold trap.

The detection of a randomly arriving muon triggers the pulsed laser system. 
With a delay of about 1\,\mus the laser pulses of 5\,mJ energy and 70\,ns pulse length are coupled through a 0.6~mm diameter hole into a multi-pass cell  formed by two elongated laser mirrors~\cite{Vogelsang:2014:Cavity} which are shaped to match the muon stopping distribution.
%
%
The laser system consists of a  thin-disk Yb:YAG oscillator-amplifier system
\cite{Antognini:2009:Disklaser,Schuhmann:2015:TDL}, a frequency doubling stage (SHG), and a Ti:Sapphire oscillator.
The Ti:Sapphire oscillator is injection seeded by a Ti:Sapphire  single-frequency cw laser
which is stabilized to a calibrated Fabry-Perot and referenced to a wavemeter with a 60\,MHz absolute accuracy.  The wavemeter was calibrated using transition lines in $^{133}$Cs and $^{83}$Kr. 
From the laser laboratory the pulses are 
sent to the target region over a distance of $\mathrm{\sim20\,m}$. An active
pointing stabilization ensures a stable coupling of the laser pulse into the optical cell. 
A $\lambda$/2 retardation plate combined with a polarizer is used to adjust from time to time  the pulse energy and maintain it roughly at 5~mJ compensating for possible drifts.
The  time and spatial distribution of the laser light in the cell is monitored using several photodiodes embedded into the mirror substrates which detect the small amount of light leaking through the mirror coatings.  

A waveform analysis was applied to the LAAPDs signals to improve energy and time resolutions and to  disentangle electrons from X-ray events yielding 
a FWHM energy resolution of 16\% at 8.2 keV~\cite{Diepold:2015:xray}.
As the experiment is taking place with only one muon at a time in the setup, the detection of the muon-decay electron after the detection of a 8~keV X-ray is used to sharpen the event selection: a reduction of the background  by more than an order of magnitude was observed when implementing this muon-decay electron cut. 
To increase the detection efficiency for  these MeV electrons   which are curling in the magnetic field, four plastic scintillators were placed radially around the He target  as shown in Fig.\,\ref{fig:setup}.

\section{Measurement and Data Analysis}
\begin{figure*}
\includegraphics[width=\columnwidth]{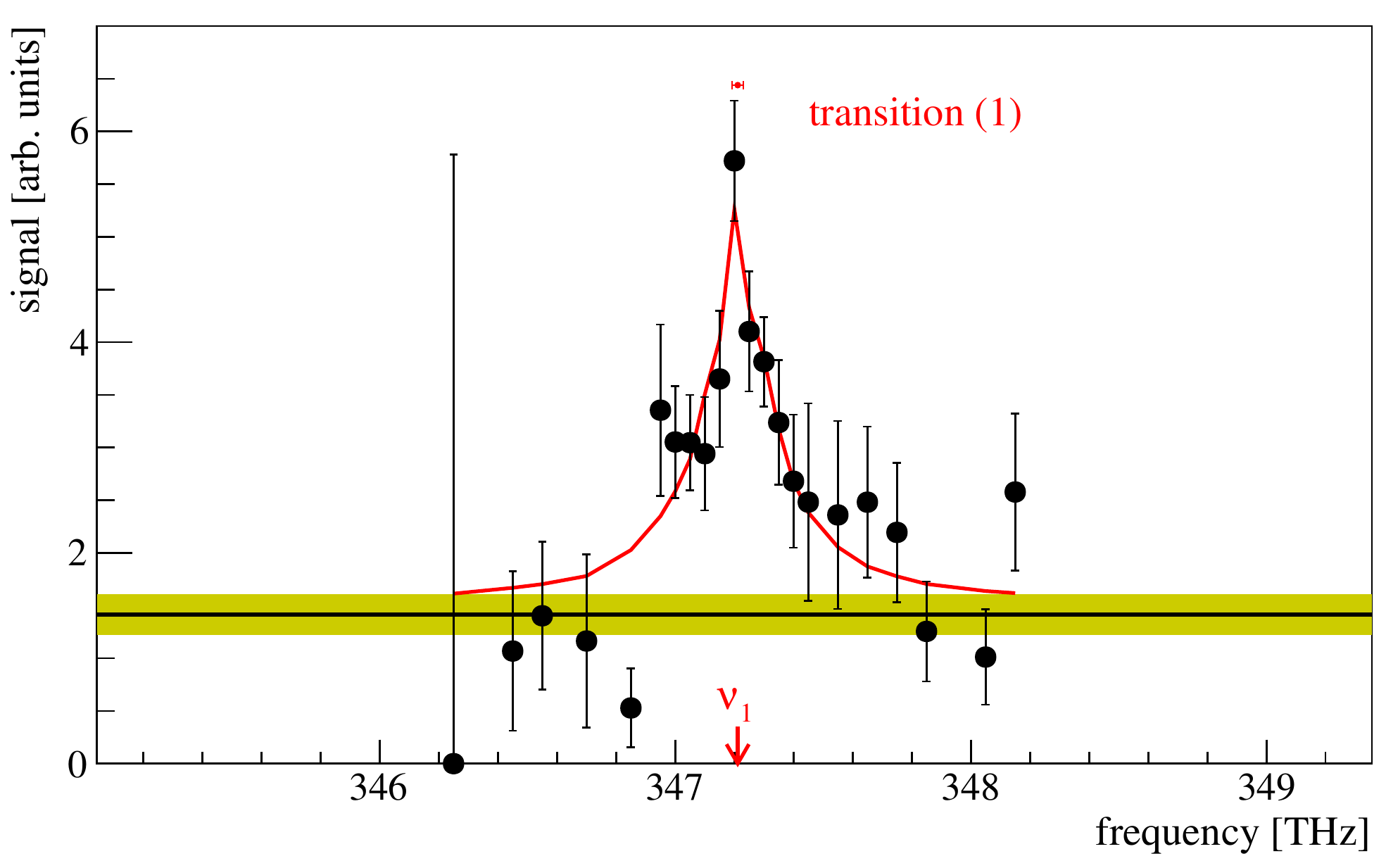}
\includegraphics[width=\columnwidth]{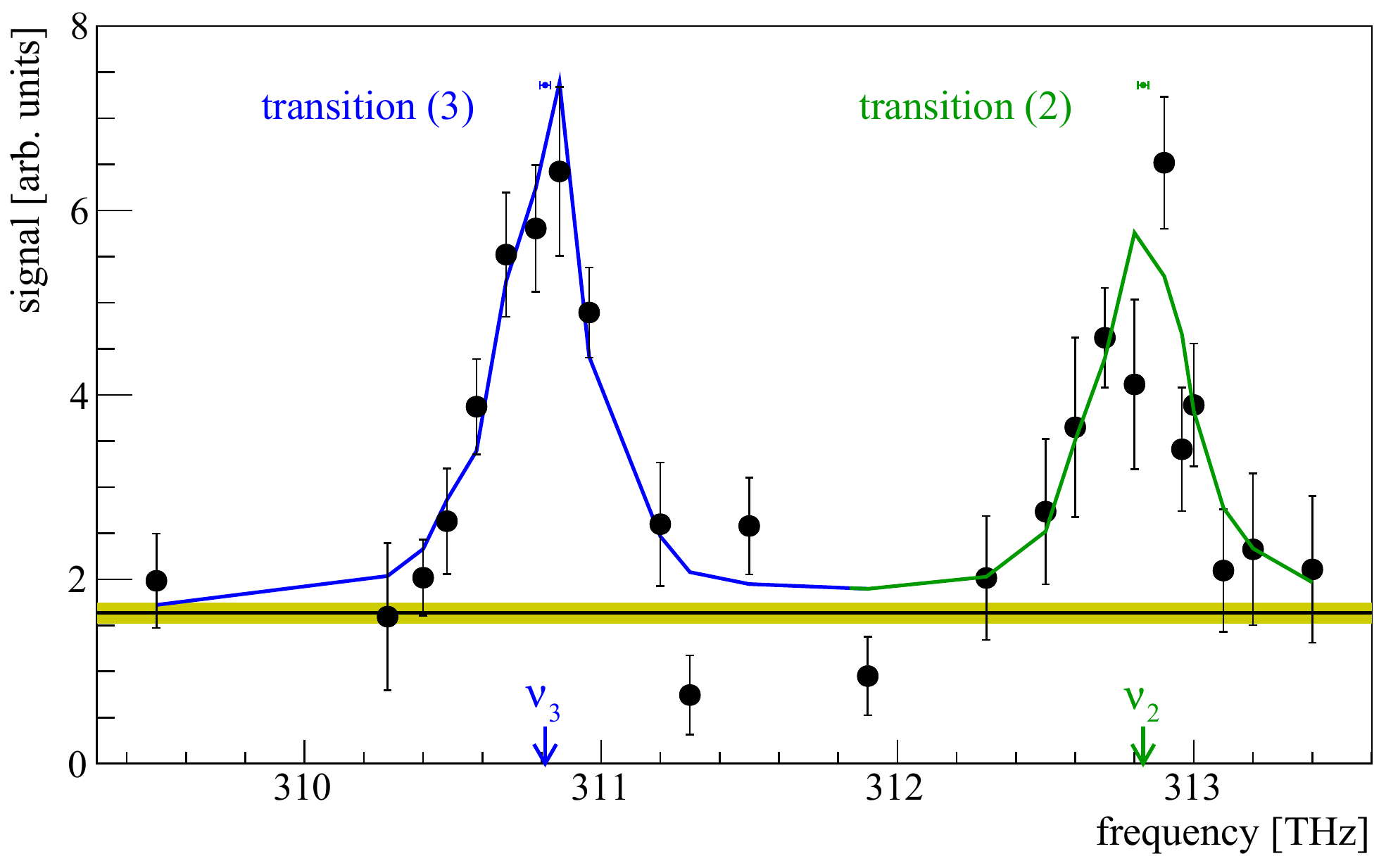}
\caption{Measured $\twoS\rightarrow\twoP$ 
  transitions in $\mu^3$He$^+$. The black data points show the number of laser-induced $K_\alpha$ X-ray events normalized to prompt $K_\alpha$ events.
   The data are fitted
  with a line shape model detailed in the main text. This model applies  only at the measured points, and the colored lines only connect the fit points. The fitted center frequencies  including their
  uncertainties are indicated by the colored points with error bar above the resonances. The yellow bands indicate the average $\pm 1 \sigma$
  backgrounds obtained from events where the laser was not fired. }
\label{fig:resonances}
\end{figure*}
The events used to expose the three 2S-2P resonances shown in Fig.~\ref{fig:levels} had to fulfill the following sequence: a muon is detected in the entrance counter, a $K_\alpha$ X-ray is detected in the LAAPDs in time coincidence with the laser light in the cavity,  and a decay electron is detected afterwards in the electron counters (or LAAPDs).  
To obtain the resonances shown in Fig.\,\ref{fig:resonances},  the number of these events  is plotted versus the laser frequency and normalized to the number of prompt $K_\alpha$ X-rays so that variations of the number of muons and X-ray detection efficiency do not affect the results.
The laser frequency  was alternated on the two sides of the resonance every 1-2 hours to reduce a possible distortion of the measured resonance due  to variations of the performance of the  experimental setup.
The injected pulse energy was limited to 5~mJ to avoid significant saturation effects and optically-induced damage of the cavity coating.
Because a minimum pulse-to-pulse separation of about 2\,ms was imposed on the laser system for increased stability, about half of the entering muons do not trigger the laser system. 
Yet these "laser-off events"  are used to precisely measure the average background yielding  the yellow bands in Fig.\,\ref{fig:resonances}.

The transition (1) took about two
weeks of continuous measurement, the transitions (2) and (3) have been measured at once, within about three weeks of continuous data taking.
For the first transition, on resonance we observed a rate
of 7-8 events/h including a background rate of $\sim$1.5 events/h.

The center frequency of the measured lines is obtained by fitting the data
with a line shape model. The model is a Lorentzian corrected for saturation effects and variations of the laser pulse energy measured for every shot.
It can therefore only be
evaluated at the position of each data point. This leads to the distortion in the line shape seen in Fig.\,\ref{fig:resonances}.
From the line shape fit we obtain following transition frequencies:
\begin{align}
\nu_{\rm exp}^{(1)} \equiv \nu(2P_{3/2}^{F=2}-2S_{1/2}^{F=1}) &= 347.212(20)^{\rm stat}(1)^{\rm sys}\thz\label{eq:freq1}\\
\nu_{\rm exp}^{(2)} \equiv \nu(2P_{3/2}^{F=1}-2S_{1/2}^{F=0})&= 312.830(21)^{\rm stat}(1)^{\rm sys}\thz\label{eq:freq2}\\
\nu_{\rm exp}^{(3)} \equiv \nu(2P_{1/2}^{F=1}-2S_{1/2}^{F=1})&= 310.814(20)^{\rm stat}(1)^{\rm sys}\thz\label{eq:freq3}.
\end{align}
The fit was done with a fixed linewidth of $\Gamma = 318.7$~GHz at FWHM. A separate fit with a free width resulted in widths that agreed with the theoretical one.
Simply fitting Lorentzians yields line centers in agreement with the ones from the line shape model.
The measured transition frequencies yield the energy splittings 
\begin{align}
\Delta E_{\rm exp}^{(1)} &= 1435.951(81)\,{\rm meV}\\
\Delta E_{\rm exp}^{(2)} &= 1293.759(86)\,{\rm meV}\\
\Delta E_{\rm exp}^{(3)} &= 1285.425(81)\,{\rm meV}
\end{align}
via the relation
$1\mev\,\widehat=\,241.798935\,\ghz$.

The uncertainty of about 20~GHz, corresponding to about $6\times 10^{-2}\,\Gamma$, is dominated by far by statistics.
The largest systematic uncertainty stems from an upper limit to
quantum interference effects~\cite{Brown:2013:QI} ($<5\times 10^{-4}\,\Gamma=0.16 $~GHz)\,\cite{Amaro:2015:muonicQI}.
The uncertainty on the laser frequency is dominated by the chirp and the calibration of the wavemeter and conservatively given by 0.1\,GHz. 
We corrected for the Zeeman shift caused by the 5T field by maximally 0.3~GHz (depending on the transition) so that the uncertainty of this correction is negligibly small in our context.
Other systematic uncertainties (light shift, collisional effects, Stark shift etc.) are negligible compared with the precision of the measurement due to the strong binding, the small atomic size and the large separation between energy levels.
\section{Experimental results}
From the three transition measurements between 2S and 2P states with different fine and hyperfine sublevels, it is possible  to determine three quantities: we choose the Lamb shift    $E_{\rm LS}=\Delta E (2P_{1/2}-2S_{1/2})$, the 2S hyperfine splitting $E_{\rm HFS}$, and the 2P fine splitting $E_{\rm FS}=\Delta E (2P_{3/2}-2P_{1/2})$.
The relations between the measured transition energies and these quantities are given by (see also Methods):
\begin{eqnarray}
\Delta E_{\rm exp}^{(1)} = E_{\rm LS} ~ - ~\frac{1}{4} E_{\rm HFS} + E_{\rm FS} -
      ~ \, 9.23945(26)\;\mathrm{meV}  \mbox{~ ~} \label{eq:trans1a}\\
\Delta E_{\rm exp}^{(2)} =E_{\rm LS}  ~ + ~\frac{3}{4} E_{\rm HFS} + E_{\rm FS} +
           15.05305(44)\;\mathrm{meV}  \mbox{~ ~} \label{eq:trans2a}\\
\Delta E_{\rm exp}^{(3)} =E_{\rm LS}  ~ - ~\frac{1}{4} E_{\rm HFS} \hspace{10mm}
          -14.80851(18)\;\mathrm{meV}  \mbox{~ ~}  \label{eq:trans3a}
\end{eqnarray}
where $E_{\rm HFS} < 0$ and the numerical values of the last terms in Eqs.~(\ref{eq:trans1a}-\ref{eq:trans3a}) arise from the 2P fine and hyperfine splittings, and include the contribution due to the mixing of the F=1 levels. 
These contributions can be calculated with great precision, because the 2P wave function  of the hydrogen-like muonic He ion has negligible overlap with the nucleus, resulting in negligible contributions from nuclear size and structure corrections. 

Note that the two most recent theory papers by Karshenboim et al.,~\cite{Karshenboim:2017jtz} and Pachucki et al.~\cite{Pachucki:2023:CODATA} use different conventions for the definition of the $2P_{1/2}$ and $2P_{3/2}$ centroids, which results in differing definitions of the Lamb shift. To obtain the constant terms in Eqs.~(\ref{eq:trans1a}-\ref{eq:trans3a}) we have used the 2P levels calculated by Karshenboim et al., and modified them to account for the different definitions, such that our final result for the Lamb shift follows the convention of Pachucki et al.\ (see Methods).

We can solve the system of equations to obtain the experimental values of the Lamb shift, the 2S HFS und the 2P fine splitting:
\begin{align}
E_{\rm LS}^{\rm exp} &= ~1258.612(~86)\mev \label{eq:exp:LS}\\
E_{\rm HFS}^{\rm exp} &= -166.485(118)\mev \label{eq:exp:HFS}\\
E_{\rm FS}^{\rm exp} &= ~~144.958(114)\mev.\label{eq:exp:FS}
\end{align}
The experimental value of the fine splitting $E_{\rm FS}^{\rm exp}$ is in excellent agreement with predictions  $E_{\rm FS}^{\rm th} = 144.979(5)\mev $~\cite{Karshenboim:2017jtz}, demonstrating consistency between our three muonic transitions measurements on the one hand, and the correctness of the theory of the 2P levels on the other.
Owing to its much smaller uncertainty  and consistency with measurements, we can use the theory value of the fine splitting to solve the system of equations  Eqs.~\eqref{eq:trans1a}-\eqref{eq:trans3a} to obtain  improved values of the Lamb shift and 2S-HFS:
\begin{align}
E_{\rm LS}^{\rm exp} &= ~1258.598(~48)^\mathrm{exp}(3)^\mathrm{theo} \mev \label{eq:exp:LS-2}\\
E_{\rm HFS}^{\rm exp} &= -166.496(104)^\mathrm{exp} (3)^\mathrm{theo}\mev \label{eq:exp:HFS-2} \, .
\end{align}
The theoretical uncertainties are from the $\pm 0.005$\,meV estimated higher-order corrections to the fine structure in Ref.~\cite{Karshenboim:2017jtz}.

\section{The helion charge radius and the isotopic shift}

\begin{figure}
  \includegraphics[width=\linewidth]{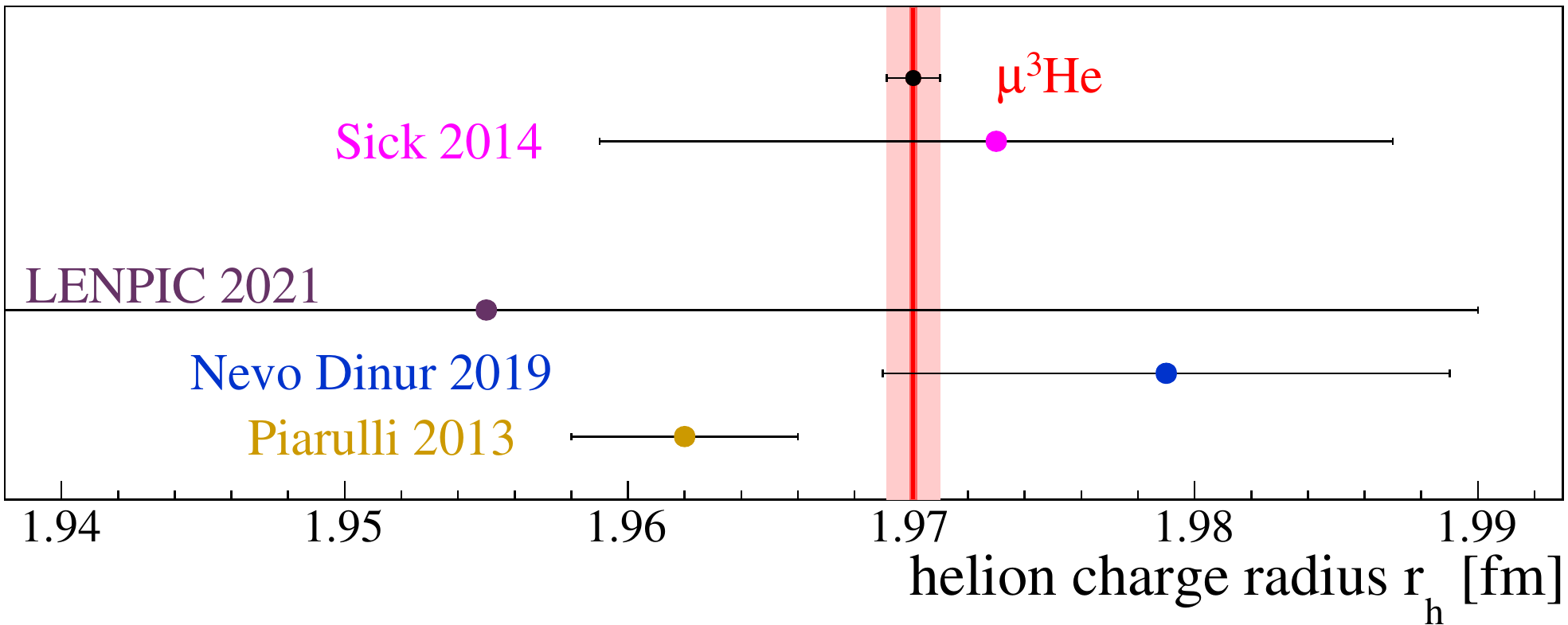}
  \caption{Recent determinations of the \Het nucleus (helion) charge radius.  The dark and light bands indicate the experimental and total 
    uncertainty in our measurement. 
    The value of Sick of 1.973(14)\,fm~\cite{Sick:2014:HeZemach}
    is from the world data on elastic electron scattering. The other values are recent predictions from nuclear few-body theory:  Piarulli 1.962(4)~fm~\cite{Piarulli:2012bn, Marcucci:2015rca},  Nevo Dinur 1.979(10)~fm~\cite{NevoDinur:2019:zemach}, and LENPIC collaboration 1.955(34)~fm~\cite{PhysRevC.103.054001, LENPIC:2022cyu}~\footnote{We obtained this  value using the point-proton structure radius and procedure explained in Ref.~\cite{PhysRevC.103.054001}.}}
  \label{fig:mu3he_radii}
\end{figure}

The theory prediction of the Lamb shift has been recently updated accounting for the  contributions of various groups. It reads~\cite{Pachucki:2023:CODATA} 
%
\begin{eqnarray}
E_{\rm LS}^\mathrm{th}(\rh^2) = & 1644.348(8)\mev - 103.383\, \rh^2\mev/\fm^2 \nonumber \\
    & +15.499(378)\mev~\mbox{~ ~ ~}
    \label{eq:theo:LS}
\end{eqnarray}
where the first term accounts for all QED corrections independent of the nuclear structure, the term proportional to \rh$^2$  accounts for the finite-size correction including radiative corrections to it,  and  the last  term is a sum of all higher-order nuclear structure dependent contributions~\cite{Carlson:2016cii, Ji:2018:nuclPOL, Pachucki:2018yxe} which are dominated by the nuclear two- and three-photon exchange contributions (2PE and 3PE, respectively).
Comparing this theory prediction with the measured Lamb shift  $E_{\rm LS}^{\rm exp}$
 we obtain the rms charge radius of the helion 
\begin{equation}
\begin{split}
\rh &= 1.97007(12)^{\rm exp}(93)^{\rm theo}\fm = 1.97007(94)\fm.
\end{split}
\end{equation}
This value is 15 times more precise than the previous best value from elastic
electron-$^3$He scattering of 1.973(14)\fm \cite{Sick:2014:HeZemach}, and in perfect agreement with it (see Fig.\,\ref{fig:mu3he_radii}).

Our value could be further improved by almost an order of magnitude by advancing the predictions for the two-photon-exchange and three-photon-exchange contributions both for the nucleus and the nucleons~\cite{Pachucki:2023:CODATA, Antognini:2022xoo, Ji:2018:nuclPOL}.

It is interesting to compare this value with the helion charge radius as 
obtained from most recent  nuclear theories which 
uses chiral effective field theory to describe the nuclear interaction and ab-initio  methods to solve the quantum-mechanical few-body problem.
Figure.~\ref{fig:mu3he_radii} shows some of the 
most recent results taken from Ref.~\cite{Marcucci:2015rca, Piarulli:2012bn, NevoDinur:2018hdo, PhysRevC.103.054001, LENPIC:2022cyu} 
depicting an overall  satisfactory agreement between the measured value and the various predictions, and  highlighting the role of the helion charge radius  as benchmark for precision nuclear theory.

\begin{figure}
  \includegraphics[width=\linewidth]{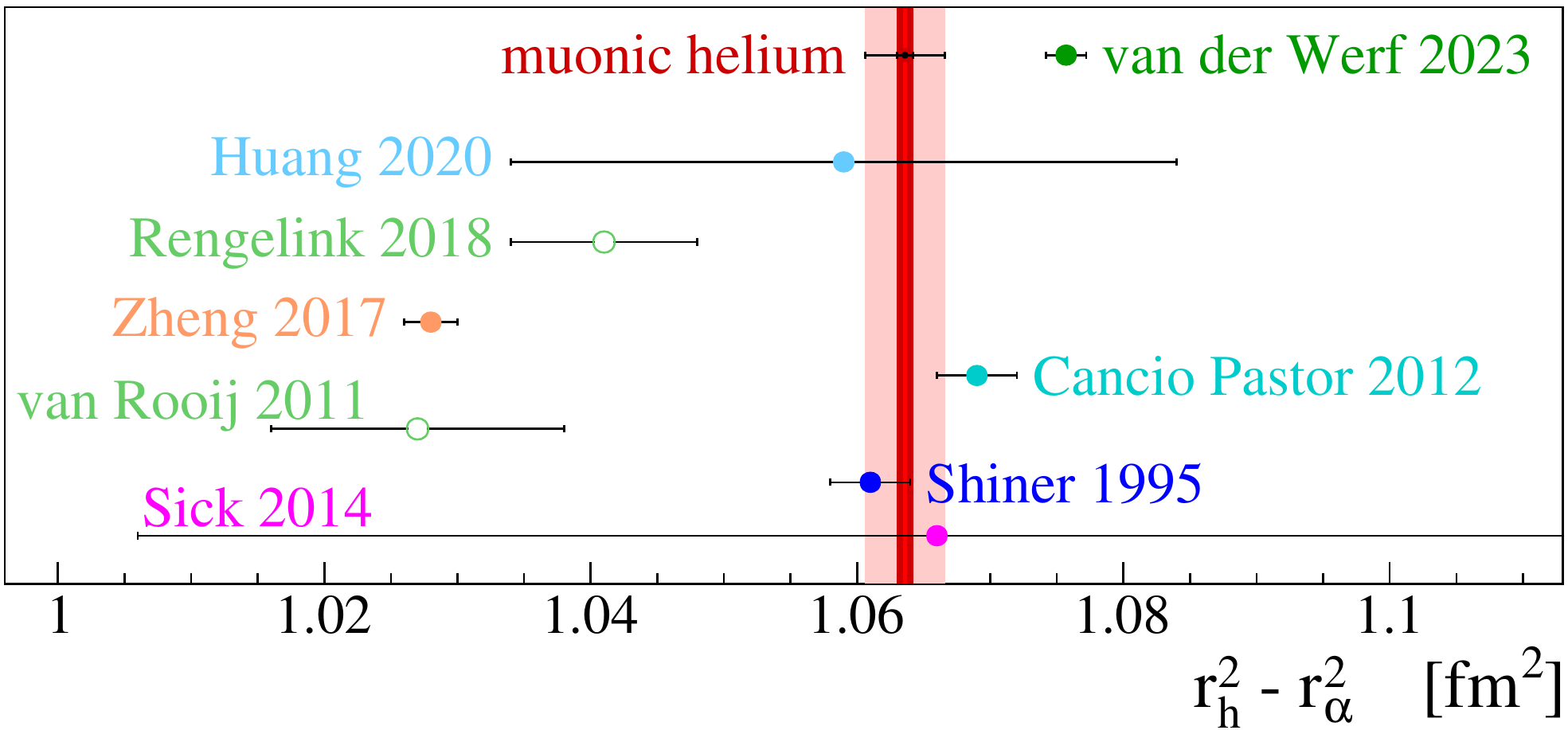}
  \caption{Squared charge radius difference $\rh^2-\ra^2$ from measurements
    of the isotope shift in ordinary He atoms
    \cite{Shiner:1995:heliumSpec,Rooij:2011:HeSpectroscopy,    CancioPastor:2012:PRL108,Patkos:2016:HeIso,Zheng:2017:He_Iso,Rengelink:2018:4He, PhysRevA.101.062507,vanderWerf:2023cuv}, compared to our value from
    muonic ions. The dark and light bands indicate the experimental and total 
    uncertainty of our determination.
    The values of 
    Shiner~\cite{Shiner:1995:heliumSpec} and Cancio
    Pastor~\cite{CancioPastor:2012:PRL108} have been corrected for improved
    theory calculations~\cite{PhysRevA.95.062510},
    but may lack a systematic correction due to 
    quantum interference effects~\cite{Brown:2013:QI}, as suggested in Ref.~\cite{Marsman:2014:QIshifts}.
    The value of Zheng~\cite{Zheng:2017:He_Iso} may have to be corrected for a systematic Doppler shift~\cite{Wen:2023:PRA107}.
    The most recent work from Amsterdam by van der Werf~\cite{vanderWerf:2023cuv} supersedes the results from van Rooij~\cite{Rooij:2011:HeSpectroscopy} and Rengelink~\cite{Rengelink:2018:4He} by the same group.
  \label{fig:he_iso}}
\end{figure}

Spectroscopy of "normal" helium atoms can  not yet provide precise values of the helion and alpha-particle charge radii, given the present uncertainty of the  three-body atomic theory.
Yet, in the isotopic shift, several cancellations take place in the theory~\cite{PhysRevA.95.062510} of the energy levels, 
so that values of $\rh^2-\ra^2$~\cite{Shiner:1995:heliumSpec,Rooij:2011:HeSpectroscopy,   CancioPastor:2012:PRL108,Patkos:2016:HeIso,Zheng:2017:He_Iso,Rengelink:2018:4He}
can be obtained where $\ra$ is the alpha particle ($^4$He) charge radius.
The scattering of the values obtained so far shown in Fig.\,\ref{fig:he_iso}  however reveals some tensions  that highlight the challenges faced by both theory and experiments.

It is thus interesting to address this quantity  by considering  the isotopic shift in muonic helium ions and presenting it in the form:~\cite{Pachucki:2023:CODATA}
%
\begin{equation}\label{eq:iso-shift}
  \rh^2-\ra^2 =-\frac{E_\mathrm{LS}^\mathrm{exp}(\mu^3\mathrm{He}^+)}{103.383\mathrm{\, \frac{meV}{fm^2}}}+\frac{E_\mathrm{LS}^\mathrm{exp}(\mu^4\mathrm{He}^+)}{106.209 \mathrm{\, \frac{meV}{fm^2}}}+ 0.2585(30)\,\mathrm{fm^2}
\end{equation}
to take advantage of some cancellations in the nuclear structure contributions.
Inserting the measured Lamb shift in $\mu^4$He$^+$ of $E_\mathrm{LS}^\mathrm{exp}(\mu^4\mathrm{He}^+)=1378.521(48)$~meV~\cite{Krauth:2021foz} and $E_\mathrm{LS}^\mathrm{exp}(\mu^3\mathrm{He}^+)$ from Eq.~\eqref{eq:exp:LS}
into this expression~\cite{Diepold:2016:mu4HeTheo,Pachucki:2023:CODATA}
we obtain 
\begin{equation}\label{eq:iso:final}
  \rh^2-\ra^2 = 1.0636(6)^{\rm exp}(30)^{\rm theo}\fm^2.
\end{equation}
This value can be compared with various isotopic shift measurements obtained in regular He atoms with two electrons~\cite{Shiner:1995:heliumSpec,Rooij:2011:HeSpectroscopy,    CancioPastor:2012:PRL108,Patkos:2016:HeIso,Zheng:2017:He_Iso,Rengelink:2018:4He, PhysRevA.101.062507}, where long-standing discrepancies exist, see Fig.~\ref{fig:he_iso}. 
Several novel systematic shifts were recently identified in these measurements~\cite{Marsman:2014:QIshifts,Wen:2023:PRA107,vanderWerf:2023cuv}, but nevertheless our result deviates by 3.6$\sigma$ from the most recent and most precise value by the Amsterdam group~\cite{vanderWerf:2023cuv}.
Our measurements in muonic helium ions do not require the exquisite experimental accuracy of $10^{-12}$ reached in normal helium atoms, and are not sensitive to systematic effects. On the other hand, our results are limited by the nuclear structure effects which are much larger in muonic systems~\cite{Ji:2018:nuclPOL,NevoDinur:2018hdo, Hernandez:2019zcm, PhysRevC.103.024313,Acharya:2022drl, Muli:2022jma}.
Clearly, more work is required to understand this $3.6\sigma$ deviation.

For completeness we quote here the updated
Lamb shift theory in $\mu^4$He$^+$ from Ref.~\cite{Pachucki:2023:CODATA} 
\begin{equation}
\begin{split}
E_{\rm LS}^\mathrm{th}(\ra^2) =& 1668.491(7)\mev - 106.209\ra^2\;\mev/\fm^2\\
    &+9.276(433) \mev \; . \label{eq:theo:LS-4He}
\end{split} 
\end{equation}
Combined with the Lamb shift in $\mu^4$He$^+$ we have measured in Ref.~\cite{Krauth:2021foz} this yields an updated $^4$He (alpha particle) charge radius
of $r_\alpha=1.6786(12)$~fm with 
a 45\% larger uncertainty compared to our previous determination which used  the theory summarized in~\cite{Diepold:2016:mu4HeTheo}.
The new \ra{} is obtained using the two-photon-exchange (TPE) contribution calculated solely from ab-initio theory~\cite{Ji:2018:nuclPOL} while previously we split it into a third-Zemach moment (Friar radius) contribution obtained from electron elastic scattering  and a polarizability contribution from few-nucleon theories~\cite{Ji:2018:nuclPOL}. Because this splitting could lead to some 
inconsistencies~\cite{Pachucki:2023:CODATA}, we now opt for the solution fully based on few-body theories~\cite{Ji:2018:nuclPOL}.

\section{Nuclear-structure contribution for the 2S HFS}

By comparing the measured 2S-HFS  $E_{\rm HFS}^{\rm exp}$ in  $\mu^3$He$^+$ with the corresponding theory prediction~\cite{Franke:2016:mu3HeTheo}
\begin{equation}
  E^{\rm th}_{\rm HFS} = -172.7457(89)\mev + E_{\rm HFS}^{\rm nucl. struct.}\label{eq:theo:HFS}
\end{equation}
we extract the nuclear structure-dependent contributions (2PE and higher orders) to the 2S hyperfine splitting
\begin{equation}
  E^{\rm nucl. struct.}_{\rm HFS} = 6.25(10)\mev,
  \label{eq:HFS-TPE}
\end{equation}
with an uncertainty arising basically only from the statistical uncertainties of the measurements. 
Subtracting the elastic part of the two-photon-exchange contribution $\Delta E_{\rm 2PE}^{\rm Zemach} = 2.5836\,{\rm meV/fm}~r_{Z} = 6.53(4)$\,meV, 
where $r_{Z}=2.528(16)$~fm is the Zemach radius of the $^3$He nucleus~\cite{Sick:2014:HeZemach, NevoDinur:2018hdo} from $E^{\rm nucl. struct.}_{\rm HFS}$, we can obtain  a value for the hitherto unknown polarizability contribution to the  \twoS{} hyperfine splitting of $-0.28(10)\mev$ (that includes also higher-order contributions).
This represents an important benchmark to refine our understanding of the magnetic structure of the \Het{} nucleus.
It also allows, using appriopiate scaling, to predict the nuclear structure contribution of the ground-state hyperfine splitting in  $\mu^3\mathrm{He}^+$ restricting considerably the range where this reasonance has to be searched for. 

\section{Conclusion and Outlook}

Laser spectroscopy of $\mu^3$He$^+$ and $\mu^4$He$^+$~\cite{Krauth:2021foz} ions provides precision values of $^3$He and $^4$He charge radii and two-photon-exchange contributions, 
that serve as benchmarks~\cite{Ekstrom:2015:PRC91,Lynn:2017:PRC,Vanasse:2018:PRC} for few-body ab-initio nuclear theories~\cite{Hammer:2020:RMP}, driving the theory to new levels of precision.
The nuclear theory~\cite{Hammer:2020:RMP} is challenged  to scrutinize its approximations~\cite{Marcucci:2015rca}, to systematically improve the nuclear interaction~\cite{Epelbaum:2019jbv, Krebs:2019aka, PhysRevLett.126.092501,Ekstrom:2015:PRC91,Lynn:2017:PRC,Vanasse:2018:PRC} and the  formalism~\cite{Ji:2018:nuclPOL, NevoDinur:2018hdo, Hernandez:2019zcm, PhysRevC.103.024313} while finding novel methods to evaluate the uncertainties~\cite{Acharya:2022drl, Muli:2022jma}.

The precise knowledge of  \rh{} 
and \ra{} reduces the uncertainty of the nuclear-structure-dependent contributions in He and He$^+$~\cite{Moreno:2023amv}.
This paves the way for bound-state QED tests for two-body (He$^+$) and three-body (He) systems to an unprecedented level of accuracy~\cite{Karshenboim:2019iuq, Yerokhin:2019:LS, Antognini:2022xoo}.
%
The Lamb shift theory in He  has undergone a spectacular advance in recent years, completing the calculation up to terms of order $\alpha^7 m$~\cite{Patkos:2021wam}.
Once the remaining discrepancies are solved there, electronic helium atoms and ions will provide more accurate charge radii~\cite{Krauth:2019:HePlus,Moreno:2023amv}, because of the smaller higher-order nuclear structure effects in these electronic systems. Using Eqs.~(\ref{eq:theo:LS}),(\ref{eq:theo:LS-4He}), our measurements in muonic He ions will then yield the most precise experimental values for the nuclear structure contributions (beyond the leading finite-size effect).
Finally, of course, comparison of results from electronic~\cite{vanderWerf:2023cuv} and muonic systems can be used to search for physics beyond the Standard Model~\cite{Antognini:2022xoo}.

\section{Methods}

To arrive at Eqs.~(\ref{eq:trans1a}-\ref{eq:trans3a}) we used the most recent calculations for the 2P level structure by Karshenboim et al.~\cite{Karshenboim:2017jtz}, and of the 2S Lamb shift by Pachucki et al.~\cite{Pachucki:2023:CODATA}.

However, these papers use a different convention for the definition of the $2P_{1/2}$ and $2P_{3/2}$ level centroids, and of the Lamb shift. 
These differences come from the treatment of the Barker-Glover (BG) and Brodsky-Parsons (BP) terms.

Both terms are not part of what Karshenboim et al.~\cite{Karshenboim:2017jtz} denote as "Lamb shift", but it is named 
"Unperturbed quantum mechanics" therein, and their sum is -0.0032\,meV (Tab.~X No.~0), which is the sum of the "BP*" and "BG*" terms listed in Tab.II, therein.

The "BG*" term of 0.1265\,meV in Ref.~\cite{Karshenboim:2017jtz} is also already included in the Lamb shift in Pachucki et al.~\cite{Pachucki:2023:CODATA}, term as "$(Z \alpha)^4$ recoil", term III.C, hence this term contributes equally to the $2P_{1/2}$-$2S_{1/2}$ energy difference in both conventions.

The "BP*" term of -0.1298\,meV, however, is not considered in Pachucki et al.~\cite{Pachucki:2023:CODATA}, because it only arises because of the hyperfine level mixing of the F=1 levels.

Thus, to make the Lamb shift of Pachucki et al.~\cite{Pachucki:2023:CODATA} compatible with the definitions of the 2P levels in Karshenboim et al.~\cite{Karshenboim:2017jtz}, we define the position of our $2P_{1/2}$ centroid as 
\begin{align}
    E(2P_{1/2}) = E(2S_{1/2}) + E_\mathrm{LS} - 0.1298\,{\rm meV} \, ,
\end{align}
where 0.1298\,meV  is the  "BP*" term in Ref.~\cite{Karshenboim:2017jtz}, Tab.II, No.~0.4.
This ensures that our measured Lamb shift $E_\mathrm{LS}$ agrees with the convention of Pachucki, while the FS and 2P-HFS splittings follow the convention of Karshenboim, where the $2P_{1/2}$ and $2P_{3/2}$ levels are the weighted average of the physical hyperfine levels.\\
This gives for the transition energies we measured

\begin{align}
\Delta E_{\rm exp}^{(1)} &= E_{\rm LS}  + E_{\rm BP*}
    - \frac{1}{4} E_{\rm HFS} 
    + E_{\rm FS}
    + \frac{3}{8} E_{\rm HFS}^{(P3/2)}  \label{eq:trans1m}\\
\Delta E_{\rm exp}^{(2)} &= E_{\rm LS}   + E_{\rm BP*}
    + \frac{3}{4}  E_{\rm HFS}
    + E_{\rm FS}
    - \frac{5}{8} E_{\rm HFS}^{(P3/2)}  \label{eq:trans2m}\\
\Delta E_{\rm exp}^{(3)} &= E_{\rm LS}   + E_{\rm BP*}
    - \frac{1}{4}E_{\rm HFS}
    \hspace{10mm}
    + \frac{1}{4} E_{\rm HFS}^{(P1/2)} \, , \label{eq:trans3m}
\end{align}
where 
$E_{\rm HFS}^{(P1/2)}=-58.7150(7)$~meV and 
$E_{\rm HFS}^{(P3/2)}=-24.2925(7)$~meV are the 
2P hyperfine level splittings~\cite{Karshenboim:2017jtz}
and $E_\mathrm{BP*} = -3/4 * 0.17302(2)$~meV (see Tab.~XIV in Ref.~\cite{Karshenboim:2017jtz}) is the above-mentioned 
"BP*" term of $-0.1298$\,meV~\cite{Karshenboim:2017jtz}.

\section{Acknowledgments}
This paper is dedicated to Ingo Sick and Wim Vassen.
We thank Sonia Bacca, Kjeld Eikema, Evgeny Epelbaum, 
Franziska Hagelstein, 
Savely Karshenboim,
Simone Li~Muli, Vadim Lensky, 
Laura Marcucci, Ulf Mei{\ss}ner,
Krzysztof Pachucki, 
Leo Simons, and Kees Steinebach for fruitful discussions.
We acknowledge support from the European Research Council (ERC) through StG.\ \#279765, and CoG. \#725039;
%
from the Deutsche Forschungsgemeinschaft (DFG, German Research Foundation) under Germany's Excellence Strategy EXC PRISMA+ (390831469),
European Union’s Horizon 2020 research and innovation programme under grant agreement STRONG – 2020 - No
824093;
from FEDER and FCT via project PTDC/FIS-NUC/0843/2012, 
%
from FCT Contract No.\ SFRH/BPD/76842/2011, 
and SFRH/BPD/92329/2013, SFRH/BD/52332/2013;
%
%
from DFG\_GR\_3172/9-1,
from SNF projects 2200021L\_138175, 200020\_159755,  200021\_165854,
200020\_197052
and 
by the ETH Femtosecond and Attosecond Science and Technology (ETH-FAST) initiatives as part of the NCCR MUST program.


\bibliographystyle{mysty1}
\bibliography{ref}

\end{document}